\documentclass[twocolumn, aps, prc, superscriptaddress, showpacs,
floatfix]{revtex4}

\usepackage{multirow}
\usepackage{epsfig}
\usepackage{amsmath}

\usepackage{subfigure}
\usepackage{latexsym}
\usepackage{feynmp}

\usepackage[normalem]{ulem}  
\usepackage[dvips]{color} 

\renewcommand\sout{\bgroup \color{red} \ULdepth=-.5ex \ULset}

\begin{document}

\title{Reduction of the $K^*$ meson abundance in heavy ion collisions}

\author{Sungtae Cho}
\affiliation{Division of Science
Education, Kangwon National University, Chuncheon 200-701, Korea}
\author{Su Houng Lee}
\affiliation{Institute of Physics and
Applied Physics, Yonsei University, Seoul 120-749, Korea}

\begin{abstract}

We study the $K^*$ meson reduction in heavy ion collisions by
focusing on the hadronic effects on the $K^*$ meson abundance. We
evaluate the absorption cross sections of the $K^*$ and $K$ meson
by light mesons in the hadronic matter, and further investigate
the variation in the meson abundances for both particles during
the hadronic stage of heavy ion collisions. We show how the
interplay between the interaction of the $K^*$ meson and kaon with
light mesons in the hadronic medium 
determines the final yield difference of the statistical
hadronization model to the experimental measurements. For the
central Au+Au collision at $\sqrt{s_{NN}}=200$ GeV, we find that
the $K^*/K$ yield ratio at chemical freeze-out decreases by $36\%$
during the expansion of the hadronic matter, resulting in the
final ratio comparable to STAR measurements of 0.23 $\pm0.05$.

\end{abstract}



\maketitle

\section{Introduction}

Relativistic heavy ion collision experiments have enabled the
production of a system of quantum chromodynamic (QCD) matter at
extreme conditions under controlled conditions
\cite{Arsene:2004fa, Back:2004je, Adams:2005dq, Adcox:2004mh,
Gyulassy:2004zy}. Due to the huge energies available in heavy ion
collisions, it is expected that a possible phase transition
predicted by Lattice QCD \cite{Gupta:2011wh} between a hadronic
matter and a system of deconfined quarks and gluons takes place,
and the quark-gluon plasma (QGP) at very high temperature is
produced at the initial stage of the collision. As a result, large
numbers of hadronic particles are produced during the quark-hadron
phase transition at later stages of heavy ion collisions.

These hadronic particles are believed to emerge at the transition
point with the information of the matter. The statistical
hadronization model has been quite success in explaining the
measured production yields of hadrons with two parameters
characterizing the chemical freeze-out point in heavy ion
collisions; the phase transition temperature and the baryon
chemical potential \cite{BraunMunzinger:1994xr,
BraunMunzinger:1995bp, BraunMunzinger:1999qy,
BraunMunzinger:2001ip}.

All particles produced at the freeze-out, however, are subject to
further interactions with other hadrons in the hadronic matter,
leading to possible deviations in the final yield of some hadrons
from the statistical model prediction. In addition to the effects
from hadronic interactions, the lifetime of hadrons as well as the
lifetime of the hadronic matter itself plays an important role in
changing the abundance of hadrons from the yield at the chemical
freeze-out.

The abundance of hadrons stable against strong decays is expected
to be changed mostly by hadronic interactions while that of
resonances will be affected by both their interactions with other
hadrons and their strong decays when the lifetime of resonances is
comparable to, or smaller than the lifespan of the hadronic stage
in heavy ion collisions. Daughter particles of resonances are
subject to re-scatter as well in the hadronic medium, making the
reconstruction of the resonances from an invariant mass analysis
difficult.

Studying the effects from the hadronic interactions on the
abundance of resonances has been suggested as one way of
confirming the scenario about a time delay between the chemical
and thermal freeze-out \cite{Torrieri:2001ue, Bleicher:2002dm},
since a sudden hadronization in heavy ion collisions would leave
no time for resonances to decay in the hadronic medium. In
particular, the $K^*$ meson has attracted lots of attentions as
its short lifetime 4 fm/c is less than the presumed lifespan of
the hadronic stage.

The effects of hadronic interactions on the yield of the $K^*$
meson have been measured in heavy ion collisions using $K^*/K$
yield ratios. Since the $K$ meson is the ground state of the $K^*$
meson, having the same valence quarks with a different mass and
relative orientation of its quark spins, the $K^*/K$ yield ratio
is considered to be independent of the freeze-out conditions when
hadronic interactions are neglected. It has been shown that
$K^*/K$ yield ratios decrease with the increasing size of the
system at the same energy \cite{Adams:2004ep, Aggarwal:2010mt}.
Compared to p+p collisions, $K^*/K$ yield ratios in Cu+Cu and
Au+Au collisions are smaller, naively implying that $K^*$ and $K$
mesons participate in re-scattering processes during the expansion
of hadronic matter, and that the hadronic effects become larger as
the size of the hadronic matter increases.

The average transverse momentum of the $K^*$ meson measured in
heavy ion collisions \cite{Adams:2004ep, Aggarwal:2010mt}, which
is higher than that of the $K^*$ meson in p+p collisions, also
supports the re-scattering scenario about $K^*$ mesons. $K^*$
mesons with low transverse momenta escape the hadronic stage later
than $K^*$ mesons with higher transverse momenta, and thus suffer
more re-scattering in the hadronic medium. As a result the
measurement of the $K^*/K$ yield ratio $0.23 \pm 0.05$
\cite{Adams:2004ep} in Au+Au collision at $\sqrt{s_{NN}}=200 $ GeV
is smaller than the statistical model expectation $0.33 \pm 0.01$
at that collision \cite{BraunMunzinger:2001ip}. However, as we
will see, the measurement of the $K^*/K$ yield ratio inconsistent
with the statistical model prediction not only confirms the
hadronic effects on the yield of the $K^*$ meson but also provides
information on the change in the properties of the hadronic matter
at freeze-out.

With these measurements of the $K^*/K$ yield ratio in mind, we
study here the hadronic effects on the $K^*$ meson by evaluating
its absorption cross sections with pions, $\rho$, $K$, and $K^*$
mesons, and furthermore investigate variations in the $K^*$ meson
abundance during the hadronic stage of heavy ion collisions by
solving a time evolution equation for the $K^*$ meson. After the
$K^*$ meson is produced at the chemical freeze-out, it interacts
mostly with light hadrons during the expansion of the hadronic
matter.  As a result, $K^*$ mesons can be absorbed by the
co-moving light mesons, or additionally produced from scattering
between them. Thus, evaluating the $K^*$ meson cross sections by
light hadrons is necessary in estimating the hadronic effects on
the $K^*$ meson abundance in heavy ion collisions. By comparing
our results with the experimental observation in heavy ion
collisions, we understand the discrepancy of the $K^*$ meson yield
between the statistical model and the experimental measurements.

As has been stated, the scattering of the $K^*$ meson daughter
particles such as kaons in the hadronic medium also contains
useful information in understanding the hadronic effects on $K^*$
mesons. Therefore, we also take into account interactions of the
kaon with light mesons during the hadronic stage of heavy ion
collisions.

To this end, we introduce effective Lagrangians for interactions
between light mesons. The effective Lagrangian methods have been
used to calculate the scattering cross sections between $J/\psi$
and hadrons in order to estimate the amount of $J/\psi$
suppression in the hadronic matter \cite{Matinyan:1998cb,
Haglin:1999xs, Lin:1999ad, Oh:2000qr}. Recently, similar
approaches have been applied to investigate the hadronic effects
on the abundance of exotic mesons such as $D_{sJ}(2317)$
\cite{Chen:2007zp} and $X(3872)$ mesons \cite{Cho:2013rpa,
Torres:2014fxa}.

This paper is organized as follows. In Sec. II, we first consider
interactions of both the $K^*$ meson and kaon with light mesons.
Then we evaluate the absorption cross sections of both mesons in
the hadronic medium using effective Lagrangians. In Sec. III we
investigate the time evolution of the $K^*$ meson abundance by
solving the kinetic equation. In Sec. IV, we argue the important
roles of the abundance ratio of $K^*$ mesons to kaons in heavy ion
collisions. Sec. IV is devoted to conclusions.

We have used throughout the work the isospin averaged mass for all
hadrons, based on experimentally measured masses
\cite{Beringer:1900zz}, e.g., $m_{K}=495.645$ MeV.

\section{Hadronic effects on $K^*$ and $K$ mesons}

We first investigate hadronic interactions of a $K^*$ meson during
the hadronic stage of heavy ion collisions. The $K^*$ meson
produced at the chemical freeze-out can be absorbed or even
produced through interactions between mostly light mesons during
the expansion of the hadronic matter. We consider here $K^*$ meson
interacting with the pions, $\rho$, $K$, and $K^*$;
$K^*\pi\rightarrow\rho K$, $K^*\rho\rightarrow \pi K$,
$K^*\bar{K}\rightarrow \rho\pi$, $K^*\bar{K}^*\rightarrow \pi\pi$,
and $K^*\bar{K}^*\rightarrow \rho\rho$. The diagrams representing
each process are shown in Fig. \ref{KvDiagrams}. We introduce the
following Lagrangians to describe the interaction between the
$K^*$ meson and other two mesons;
\begin{eqnarray}
{\cal L}_{\pi KK^*}&=&ig_{\pi K^*K}K^{*\mu}\vec\tau\cdot(\bar K
\partial_\mu\vec\pi-\partial_\mu\bar K\vec\pi) + {\rm H.c.},
\nonumber \\
{\cal L}_{\rho KK}&=&ig_{\rho KK}(K\vec\tau\partial_\mu\bar K
-\partial_\mu K\vec\tau\bar K)\cdot\vec\rho^\mu,
\nonumber \\
{\cal L}_{\rho K^*K^*}&=&ig_{\rho K^*K^*}~[ (\partial_\mu K^{*\nu}
\vec\tau\bar{K}^*_\nu-K^{*\nu}\vec\tau\partial_\mu\bar{K^*_\nu})
\cdot\vec\rho^\mu \nonumber \\
&+& (K^{*\nu}\vec\tau\cdot\partial_\mu\vec\rho_\nu-\partial_\mu
K^{*\nu}\vec\tau\cdot\vec\rho_\nu)\bar{K}^{*\mu} \nonumber \\
&+& K^{*\mu}(\vec\tau\cdot\vec\rho^\nu\partial_\mu\bar{K^*_\nu}
-\vec\tau\cdot\partial_\mu\vec\rho^\nu \bar{K}^*_\nu) ],
\label{fLagrangians}
\end{eqnarray}
obtained from free pseudoscalar and vector meson Lagrangians by
introducing the minimal substitution. In Eq. (\ref{fLagrangians}),
$K\equiv (K^0,K^+)$ and $K^*\equiv (K^{*0},K^{*+})$ denote
strangeness pseudoscalar and vector meson doublets, respectively,
and $\vec \pi$ and $\vec \rho$ denote the pion and $\rho$ meson
isospin triplets, respectively, with Pauli matrices $\vec \tau$.
$g_{\pi K^*K}$, $g_{\rho KK}$, and $g_{\rho K^*K^*}$ are strong
coupling constants, for which we use the empirical values, $g_{\pi
K^*K}=3.25$, $g_{\rho KK}=3.05$ \cite{Brown:1991ig}. We apply the
SU(3) flavor symmetry to obtain $g_{\rho K^*K^*}=g_{\pi
K^*K}=3.25$.

Using the above interaction Lagrangians we evaluate the amplitudes
for all processes shown in Fig. \ref{KvDiagrams}. The amplitudes
of the $K^*$ meson absorption by pions, $\rho$, $K$, and $K^*$
mesons, without isospin factors and before summing and averaging
over external spins, are represented by
{\allowdisplaybreaks
\begin{eqnarray}
{\cal M}_{K^*\pi\rightarrow\rho K} &\equiv& {\cal
M}^{(a)}_{K^*}+{\cal M}^{(b)}_{K^*} \nonumber \\
{\cal M}_{K^*\rho\rightarrow \pi K} &\equiv& {\cal
M}^{(c)}_{K^*}+{\cal M}^{(d)}_{K^*} \nonumber \\
{\cal M}_{K^*\bar{K}\rightarrow \rho\pi} &\equiv& {\cal
M}^{(e)}_{K^*}+{\cal M}^{(f)}_{K^*} \nonumber \\
{\cal M}_{K^*\bar{K}^*\rightarrow \pi\pi} &\equiv& {\cal
M}^{(g)}_{K^*}+{\cal M}^{(h)}_{K^*} \nonumber \\
{\cal M}_{K^*\bar{K}^*\rightarrow \rho\rho} &\equiv& {\cal
M}^{(i)}_{K^*}+{\cal M}^{(j)}_{K^*} \label{Kvamplitudes}
\end{eqnarray} }
where the amplitudes for the first $K^*\pi\rightarrow\rho K$ and
the second process $K^*\rho\rightarrow\pi K$ are
\begin{widetext}
{\allowdisplaybreaks
\begin{eqnarray}
&&{\cal M}^{(a)}_{K^*}=g_{\pi K^*K}g_{\rho K^*K^*}\epsilon^{
\alpha}_{1}\epsilon^{*\beta}_{3}\frac{1}{t-m_{K^*}^2+im_{K^*}
\Gamma_{K^*}}\Big(-g^{\mu\nu}+\frac{(p_1-p_3)^\mu(p_1-p_3)^{\nu}}
{m_{K^*}^2}\Big)(p_2+p_4)_\mu \nonumber \\
&& \qquad\times\Big((2p_1-p_3)_\beta g_{\alpha\nu}-(p_1+p_3)_{\nu}
g_{\alpha\beta}-(p_1-2p_3)_\alpha g_{\beta\nu}\Big),
\nonumber \\
&&{\cal M}^{(b)}_{K^*}=-g_{\pi K^*K}g_{\rho KK}
\epsilon^{\mu}_{1}\epsilon^{*\nu}_{3}\frac{1}{s-m_K^2}
(p_1+2p_2)_\mu (p_3+2p_4)_\nu, \label{Matx_Kvpi2toKrho}
\end{eqnarray} }
and
\begin{eqnarray}
&&{\cal M}^{(c)}_{K^*}=-g_{\pi K^*K}g_{\rho KK}
\epsilon^{\mu}_{1}\epsilon^{\nu}_{2}\frac{1}{t-m_K^2}
(p_1-2p_3)_\mu (2p_4-p_2)_\nu, \nonumber \\
&&{\cal M}^{(d)}_{K^*}=-g_{\pi K^*K}g_{\rho K^*K^*}\epsilon^{
\alpha}_{1}\epsilon^{\beta}_{2}\frac{1}{s-m_{K^*}^2+im_{K^*}
\Gamma_{K^*}} \Big(-g^{\mu\nu}+\frac{(p_1+p_2)^\mu(p_1+p_2)^{\nu}}
{m_{K^*}^2}\Big)(p_3-p_4)_\mu \nonumber \\
&& \qquad\times\Big((2p_1+p_2)_\beta g_{\alpha\nu}-(p_1-p_2)_{\nu}
g_{\alpha\beta}-(p_1+2p_2)_\alpha g_{\beta\nu}\Big),
\label{Matx_Kvrho2toKpi}
\end{eqnarray}
respectively. Similarly, amplitudes for processes
$K^*\bar{K}\rightarrow \rho\pi$ and $K^*\bar{K^*}\rightarrow
\pi\pi$ are


\begin{figure}[t]
\begin{center}
\subfigure[]{
\includegraphics[width=0.17\textwidth]{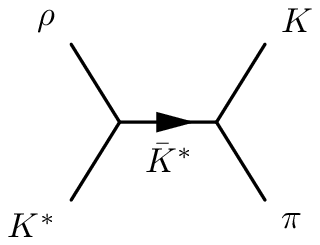}}
\quad \subfigure[]{
\includegraphics[width=0.17\textwidth]{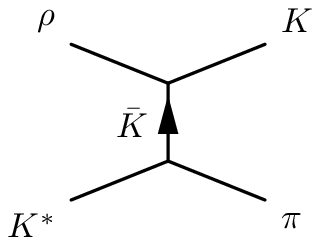}}
\quad \subfigure[]{
\includegraphics[width=0.17\textwidth]{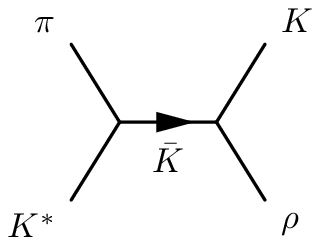}}
\quad \subfigure[]{
\includegraphics[width=0.17\textwidth]{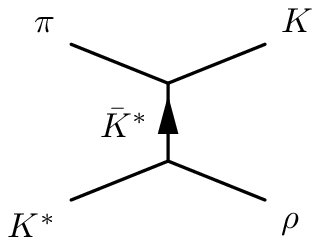}}
\quad \subfigure[]{
\includegraphics[width=0.17\textwidth]{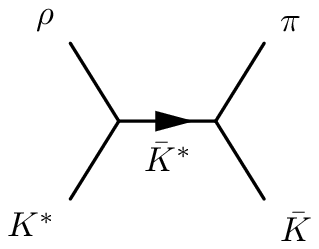}} \\
\subfigure[]{
\includegraphics[width=0.17\textwidth]{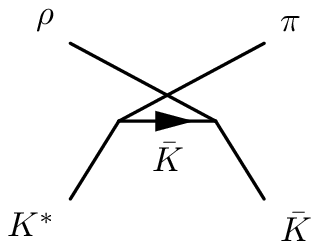}}
\quad \subfigure[]{
\includegraphics[width=0.17\textwidth]{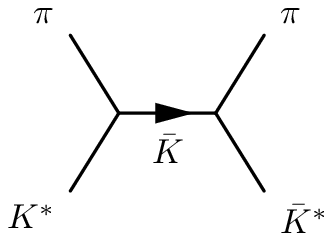}}
\quad \subfigure[]{
\includegraphics[width=0.17\textwidth]{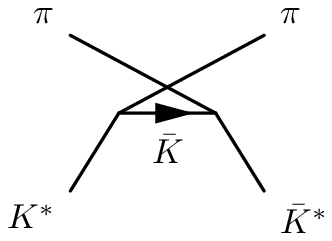}}
\quad \subfigure[]{
\includegraphics[width=0.17\textwidth]{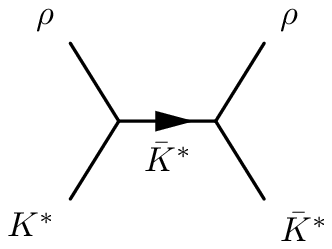}}
\quad \subfigure[]{
\includegraphics[width=0.17\textwidth]{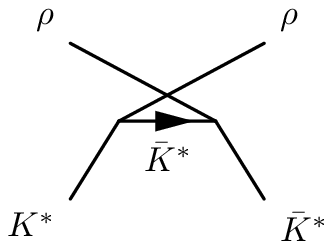}} \\
\end{center}
\caption{Born diagrams for the $K^*$ meson absorption by pions,
$\rho$, $K$, and $K^*$ mesons. $K^*\pi\to \rho K$, (a) and (b);
$K^*\rho\to \pi K$, (c) and (d); $K^*\bar{K}\to \rho\pi$, (e) and
(f); $K^*\bar{K}^*\to \pi\pi$,  (g) and (h); $K^*\bar{K}^*\to
\rho\rho$, (i) and (j). } \label{KvDiagrams}
\end{figure}

\begin{eqnarray}
&&{\cal M}^{(e)}_{K^*}=g_{\pi K^*K}g_{\rho K^*K^*}\epsilon^{
\alpha}_{1}\epsilon^{*\beta}_{3}\frac{1}{t-m_{K^*}^2+im_{K^*}
\Gamma_{K^*}} \Big(-g^{\mu\nu}+\frac{(p_1-p_3)^\mu(p_1-p_3)^{\nu}}
{m_{K^*}^2}\Big)(p_2+p_4)_\mu \nonumber \\
&& \qquad\times\Big((2p_3-p_1)_\alpha g_{\beta\nu}-(p_1+p_3)_{\nu}
g_{\alpha\beta}+(2p_1-p_3)_\beta g_{\alpha\nu}\Big), \nonumber \\
&&{\cal M}^{(f)}_{K^*}=g_{\pi K^*K}g_{\rho KK}
\epsilon^{\mu}_{1}\epsilon^{*\nu}_{3}\frac{1}{u-m_K^2}
(2p_4-p_1)_\mu (2p_2-p_3)_\nu, \label{Matx_KvKstorhopi}
\end{eqnarray}
and
\begin{eqnarray}
&&{\cal M}^{(g)}_{K^*}=g_{\pi K^*K}^2\epsilon^{\mu}_{1}\epsilon^{
\nu}_{2}\frac{1}{t-m_K^2}(p_1-2p_3)_\mu(p_2-2p_4)_\nu, \nonumber \\
&&{\cal M}^{(h)}_{K^*}=g_{\pi
K^*K}^2\epsilon^{\mu}_{1}\epsilon^{\nu}_{2}\frac{1}{u-m_K^2}
(p_1-2p_4)_\mu(p_2-2p_3)_\nu, \label{Matx_KvKv2topipi}
\end{eqnarray}
respectively. Finally, the amplitudes for $K^*\bar{K^*}\rightarrow
\rho\rho$ are

\begin{eqnarray}
&&{\cal M}^{(i)}_{K^*}=g_{\rho K^*K^*}^2\epsilon^{\alpha}_{1}
\epsilon^{*\beta}_{3}\epsilon^{\gamma}_{2}\epsilon^{*\delta}_{4}
\frac{1}{t-m_{K^*}^2+im_{K^*}\Gamma_{K^*}}
\Big(-g^{\mu\nu}+\frac{(p_1-p_3)^\mu(p_1-p_3)^{\nu}}
{m_{K^*}^2}\Big) \nonumber \\
&& \qquad\times\Big((2p_3-p_1)_\alpha g_{\beta\mu}-(p_1+p_3)_{\mu}
g_{\alpha\beta}+(2p_1-p_3)_\beta g_{\alpha\nu}\Big)
\Big((p_2+p_4)_\gamma g_{\delta\nu}+(p_2-2p_4)_{\nu}
g_{\gamma\delta}+(p_4-2p_2)_\delta g_{\gamma\nu}\Big), \nonumber \\
&&{\cal M}^{(j)}_{K^*}=g_{\rho K^*K^*}^2\epsilon^{\alpha}_{1}
\epsilon^{*\beta}_{4}\epsilon^{\gamma}_{2}\epsilon^{*\delta}_{3}
\frac{1}{u-m_{K^*}^2+im_{K^*}\Gamma_{K^*}}
\Big(-g^{\mu\nu}+\frac{(p_1-p_4)^\mu(p_1-p_4)^{\nu}}
{m_{K^*}^2}\Big) \nonumber \\
&& \qquad\times\Big((2p_4-p_1)_\alpha g_{\beta\mu}-(p_1+p_4)_{\mu}
g_{\alpha\beta}+(2p_1-p_4)_\beta g_{\alpha\nu}\Big)
\Big((p_2+p_3)_\gamma g_{\delta\nu}+(p_2-2p_3)_{\nu}
g_{\gamma\delta}+(p_3-2p_2)_\delta g_{\gamma\nu}\Big).
\label{Matx_KvKv2torhorho}
\end{eqnarray} 

\end{widetext}
In the above equations, $p_i$ denotes the momentum of particle
$i$. We keep the convention that particles $1$ and $2$ stand for
initial-state mesons, and particles $3$ and $4$ final-state mesons
on the left and right sides of the diagrams, respectively. The
Mandelstam variables $s=(p_{1}+p_{2})^{2}$, $t=(p_{1}-p_{3})^{2}
$, and $u=(p_{1}-p_{4})^{2}$ have also been used. We apply here
the $K^*$ meson propagator with its decay width, $\Gamma_{K^*}$,
and use the isospin averaged value for the $K^*$ meson decay
width, $\Gamma_{K^*}=49.1$ MeV \cite{Beringer:1900zz}.

In order to take the finite size of the hadron into consideration
when evaluating amplitudes, we apply the following form factor at
each interaction vertex for the $u,t$-channel and the $s$-channel,
respectively,
\begin{equation}
F_{u,t}(\vec q) =\frac{\Lambda^2-m_{ex}^2}{\Lambda^2 +{\vec q}^2},
\qquad F_{s}(\vec q) =\frac{\Lambda^2+m_{ex}^2}{\Lambda^2
+\omega^2}, \label{form}
\end{equation}
with ${\vec q}^2$ being the squared three-momentum transfer for
$t$ and $u$ channels, and $\omega^2$ being the total energy of the
incoming particles for $s$ channel taken in the center of mass
frame. $m_{ex}$ is the mass of the exchanged particle in each
diagram shown in Fig. \ref{KvDiagrams}. We set the cutoff
parameter $\Lambda$ to be $\Lambda=1.8$ GeV \cite{Brown:1991ig}.
The final isospin- and spin-averaged cross section is given by,
\begin{equation}
\sigma=\frac{1}{64\pi^2 s g_1g_2}\frac{|\vec p_f|}{|\vec p_i|}\int
d\Omega\overline{|\mathcal{M}|^2}F^4, \label{sigma}
\end{equation}
where $g_1$ and $g_2$ are the degeneracy factors of the initial 1
and 2 particles; $g_1=(2I_1+1)(2S_1+1)$ and
$g_2=(2I_2+1)(2S_2+1)$, respectively. $\overline{|\mathcal{M}|^2}$
represents the squared amplitude of all processes in Eq.
(\ref{Kvamplitudes}) obtained by summing over the isospins and
spins of both the initial and final particles. $|\vec p_i|$ and
$|\vec p_f|$ in Eq. (\ref{sigma}) stand for the three-momenta of
the initial and final particles in the center-of-mass frame.

\begin{figure}[t]
\begin{center}
\subfigure[]{
\includegraphics[width=0.17\textwidth]{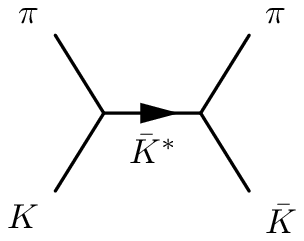}}
\quad \subfigure[]{
\includegraphics[width=0.17\textwidth]{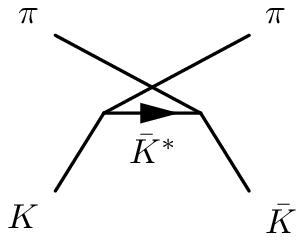}} \\
\quad \subfigure[]{
\includegraphics[width=0.17\textwidth]{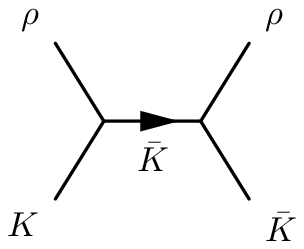}}
\quad \subfigure[]{
\includegraphics[width=0.17\textwidth]{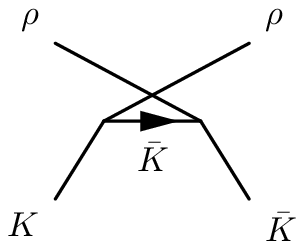}}
\end{center}
\caption{Born diagrams for the $K$ meson absorption by $\bar{K}$
mesons: $K\bar{K}\to\pi\pi$, (a) and (b); $K\bar{K}\to\rho\rho$,
(c) and (d). } \label{KDiagrams}
\end{figure}

Using the same method, we investigate hadronic effects on a $K$
meson during the hadronic stage in heavy ion collisions. We
consider interactions of the $K$ meson with pions, $\rho$, $K$,
and $K^*$ mesons; $K\pi\rightarrow\rho K^*$, $K\rho\rightarrow \pi
K^*$, $K\bar{K}\rightarrow \pi\pi$, $K\bar{K}\rightarrow
\rho\rho$, and $K\bar{K}^*\rightarrow \pi\rho$. Among these,
however, two processes $K\pi\rightarrow\rho K^*$,
$K\rho\rightarrow \pi K^*$, are same as the inverse processes of
the $K^*$ meson interacting with $\rho$ mesons and pions as shown
in Fig. \ref{KvDiagrams}; (c) and (d), (a) and (b), respectively,
and the process $K\bar{K}^*\rightarrow \pi\rho$ is same as that of
the $K^*$ meson interacting with $\bar{K}$ mesons, diagrams (e)
and (f) shown in Fig. \ref{KvDiagrams}. Therefore, all we need to
consider more are the following amplitudes,
\begin{eqnarray}
{\cal M}_{\bar{K}K\rightarrow\pi\pi} &\equiv& {\cal
M}^{(a)}_K+{\cal M}^{(b)}_K, \nonumber \\
{\cal M}_{\bar{K}K\rightarrow\rho\rho} &\equiv& {\cal
M}^{(c)}_K+{\cal M}^{(d)}_K, \label{amplitudes}
\end{eqnarray}
for processes $K\bar{K}\rightarrow \pi\pi$ and
$K\bar{K}\rightarrow \rho\rho$. We show the diagrams for theses
processes in Fig. \ref{KDiagrams}.

The amplitudes for processes $K\bar{K}\rightarrow \pi\pi$ and
$K\bar{K}\rightarrow \rho\rho$ are {\allowdisplaybreaks
\begin{eqnarray}
&&{\cal M}^{(a)}_K=\frac{g_{\pi K^*K}^2}{t-m_{K^*}^2+im_{K^*}
\Gamma_{K^*}}(p_1+p_3)_\mu(p_2+p_4)_\nu \nonumber \\
&&\qquad \times
\Big(-g^{\mu\nu}+\frac{(p_1-p_3)^\mu(p_1-p_3)^{\nu}}
{m_{K^*}^2}\Big), \nonumber \\
&&{\cal M}^{(b)}_K=\frac{g_{\pi K^*K}^2}{u-m_{K^*}^2+im_{K^*}
\Gamma_{K^*}}(p_1+p_4)_\mu(p_2+p_2)_\nu \nonumber \\
&&\qquad \times \Big(-g^{\mu\nu}+\frac{(p_1-p_4)^\mu
(p_1-p_4)^{\nu}}{m_{K^*}^2}\Big), \label{Matx_KsKstopipi}
\end{eqnarray}}
and
\begin{eqnarray}
&&{\cal M}^{(c)}_K=g_{\rho
KK}^2\epsilon^{*\mu}_{3}\epsilon^{*\nu}_{4}
\frac{1}{t-m_K^2} (2p_1-p_3)_\mu(2p_2-p_4)_\nu, \nonumber \\
&&{\cal M}^{(d)}_K=g_{\rho KK}^2\epsilon^{*\mu}_{4}
\epsilon^{*\nu}_{3} \frac{1}{u-m_K^2}(2p_1-p_4)_\mu(2p_2-p_3)_\nu,
\nonumber \\ \label{Matx_KsKstorhorho}
\end{eqnarray}
respectively. Then, using Eqs. (\ref{form}) and (\ref{sigma}) we
evaluate the $K$ meson absorption cross sections.

Lastly we consider the possibility of the $K^*$ meson formation
from pions and kaons. The scattering cross section for the $K^*$
meson production is given by the spin-averaged relativistic
Breit-Wigner cross section;
\begin{equation}
\sigma_{K\pi\to K^*}=\frac{g_{K^*}}{g_K g_{\pi}}
\frac{4\pi}{p_{cm}^2}\frac{s\Gamma_{K^*\to
K\pi}^2}{(m_{K^*}-\sqrt{s})^2+s\Gamma_{K^*\to K\pi}^2},
\label{BWsigma}
\end{equation}
with $g_\pi$, $g_K$, and $g_{K^*}$ being the degeneracy of pions,
$K$, and $K^*$ mesons, $g_i=(2S_i+1)(2I_i+1)$, respectively, and
$p_{cm}$ the momentum in the center-of-mass frame.
$\Gamma_{K^*\to K\pi}$ is the total decay width for a reaction
$K\pi\to K^*\to K\pi$ as a function of $\sqrt{s}$. We take the
following $\sqrt{s}$-dependent decay width $\Gamma_{K^*\to K\pi}$
of the $K^*$ meson;
\begin{equation}
\Gamma_{K^*\to K\pi}(\sqrt{s})=\frac{g_{\pi K^*K}^2}{2\pi s}
p_{cm}^3(\sqrt{s}), \label{totalG}
\end{equation}

\begin{figure}[!h]
\begin{center}
\includegraphics[width=0.52\textwidth]{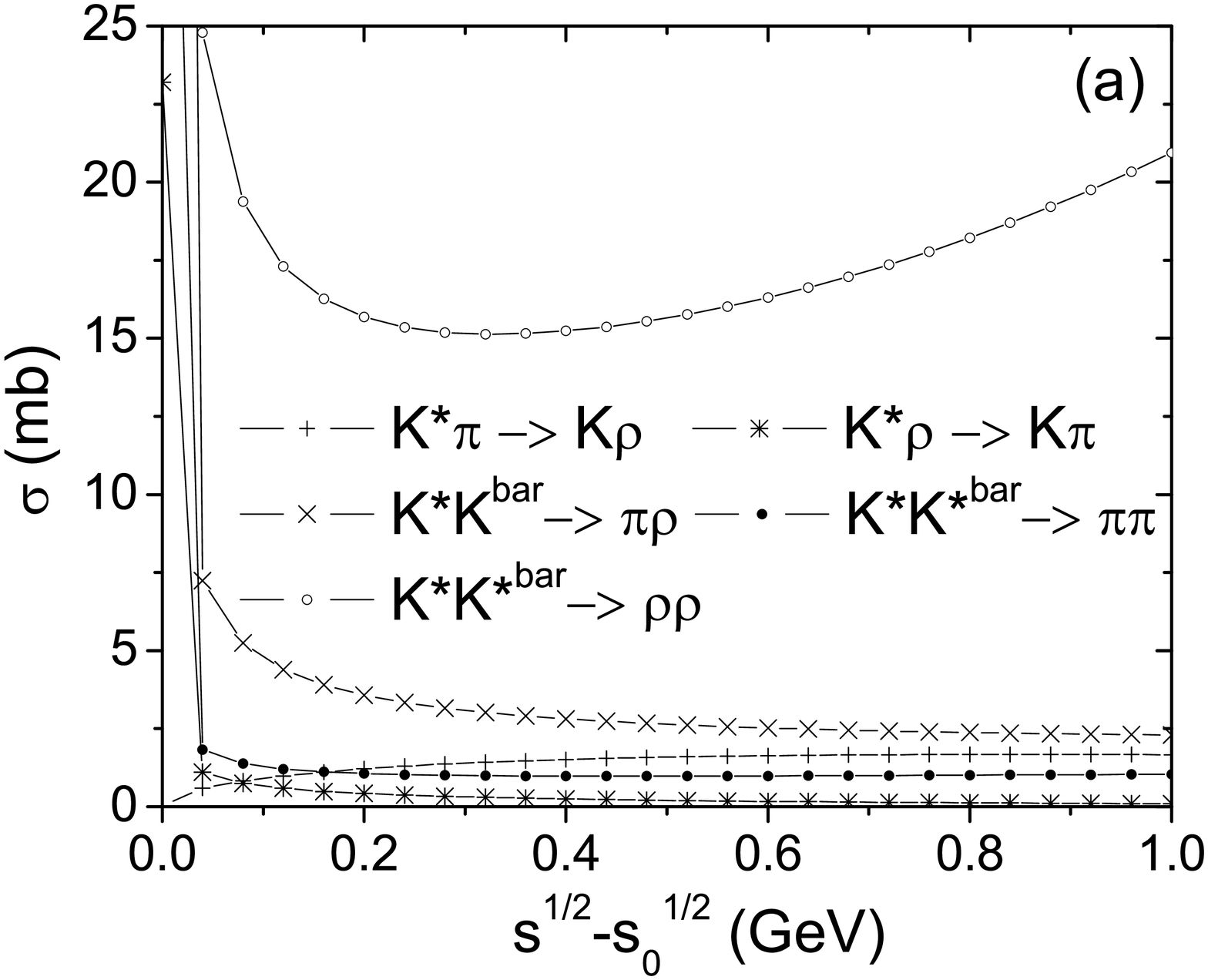}
\includegraphics[width=0.54\textwidth]{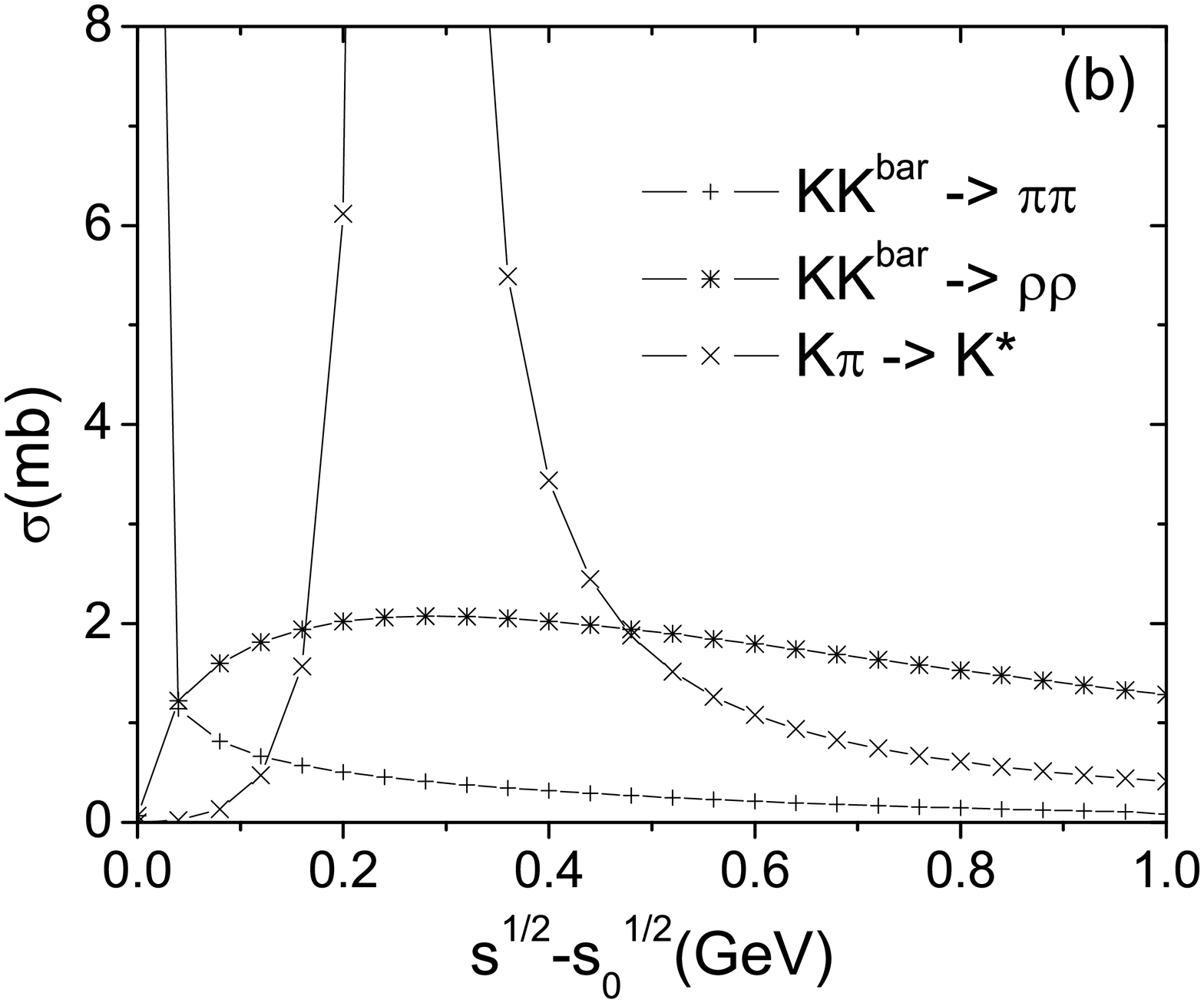} \\
\end{center}
\caption{Absorption cross sections for (a) the $K^*$ meson by
pions, $\rho$, $K$, and $K^*$ mesons via processes
$K^*\pi\rightarrow\rho K$, $K^*\rho\rightarrow \pi K$,
$K^*\bar{K}\rightarrow \rho\pi$, $K^*\bar{K}^*\rightarrow \pi\pi$,
and $K^*\bar{K}^*\rightarrow \rho\rho$, and those for (b) the $K$
meson via processes $K\bar{K}\rightarrow \pi\pi$,
$K\bar{K}\rightarrow \rho\rho$, and $K\pi\rightarrow K^*$. }
\label{KvKssigma}
\end{figure}
We show in Fig. \ref{KvKssigma} the cross sections for the
absorption of both the $K^*$ meson and the $K$ meson by pions,
$\rho$, $K$, and $K^*$ mesons via processes shown in Fig.
\ref{KvDiagrams} and Fig. \ref{KDiagrams} as functions of the
total center-of-mass energy $s^{1/2}$ above the threshold energy
$s_0^{1/2}$ of each process. We see in Fig. \ref{KvKssigma} the
general pattern that the cross sections have a peak near the
threshold energy for the endothermic processes, e.g.,
$K^*\pi\rightarrow\rho K$ and $K\bar{K}\rightarrow \rho\rho$,
while the cross sections for the exothermic processes, e.g.,
$K^*\rho\rightarrow \pi K$, $K^*\bar{K}\rightarrow \rho\pi$,
$K^*\bar{K}^*\rightarrow \pi\pi$, $K^*\bar{K}^*\rightarrow
\rho\rho$, and $K\bar{K}\rightarrow \pi\pi$, become infinite near
the threshold. However, two endothermic processes considered here,
$K^*\pi\rightarrow\rho K$ and $K\bar{K}\rightarrow \rho\rho$ show
very broad peaks above the threshold energy.

We notice that $K^*$ meson is absorbed more by pions than by
$\rho$ mesons; the absorption cross section of the $K^*$ meson by
pions, $K^*\pi\rightarrow\rho K$ is larger than that by $\rho$
mesons $K^*\rho\rightarrow \pi K$. We also see that the
annihilation cross sections for both the $K^*$ meson and $K$ meson
are larger when $\rho$ mesons are produced than when pions are
produced; the cross section for $K^*\bar{K}^*\rightarrow \rho\rho$
is larger than that for $K^*\bar{K}^*\rightarrow \pi\pi$, and the
cross section for $K\bar{K}\rightarrow \rho\rho$ is also larger
than that for $K\bar{K}\rightarrow \pi\pi$.

The cross section for $K^*\bar{K}^*\rightarrow \rho\rho$ is an
order of magnitude larger than other processes, and seems to
reflect the effect from two interaction mechanisms between three
vector mesons. All particles participating in the process
$K^*\bar{K}^*\rightarrow \rho\rho$ are vector mesons, and thus two
$\mathcal{L}_{\rho K^*K^*}$ in Eq. (\ref{fLagrangians}) are needed
to describe the process $K^*\bar{K}^*\rightarrow \rho\rho$. It has
already been shown that the interaction between three vector
mesons increases the absorption cross section in the effective
Lagrangian approach \cite{Lin:1999ad}.

It seems awkward to observe that the cross section for
$K^*\bar{K}^*\rightarrow \rho\rho$ should rise with increasing
energy even though the form factor has been correctly used to kill
the artificial growth of the cross section with the energy. This
behavior reminds the rise of the total cross section for
$p\bar{p}$ collisions at high energy. It has been already well
known that the resonance exchange is largely responsible for an
increase of the cross section in high energy scattering. In this
study the $K^*$ meson exchange in the reaction
$K^*\bar{K}^*\rightarrow \rho\rho$ causes the rise of cross
section even at relatively low energy less than 1 GeV. The
introduction of the decay width in the propagator $\Gamma_{K^*}$,
however, does not contribute to this behavior at all. Instead it
merely reduces a little bit the amplitudes for process having a
$K^*$ meson exchange. Finally, we also find that the cross section
for the formation of the $K^*$ meson from pions and $K$ mesons,
Eq. (\ref{BWsigma}) is not small at all, compared to cross
sections for other processes.

\section{Time evolutions of the $K^*$ and $K$ meson abundances}

We consider the time evolutions of the abundance for both the
$K^*$ meson and kaon based on the cross sections evaluated in the
previous section. We build a coupled evolution equation for both
particles consisting of densities and abundances for mesons
participating in all processes shown in Fig. \ref{KvDiagrams};
pions, $\rho$, $K$, and $K^*$ mesons.
\begin{widetext}
{\allowdisplaybreaks \begin{eqnarray}
\frac{dN_{K^*}(\tau)}{d\tau}&=&\langle\sigma_{K\rho\to
K^*\pi}v_{K\rho} \rangle n_{\rho}(\tau)N_{K}(\tau)-\langle
\sigma_{K^*\pi\to K\rho}v_{K^*\pi}\rangle n_{\pi}(\tau)
N_{K^*}(\tau)  \nonumber \\
&+&\langle\sigma_{K\pi\to K^*\rho}v_{K\pi} \rangle
n_{\pi}(\tau)N_{K}(\tau)-\langle \sigma_{K^*\rho\to
K\pi}v_{K^*\rho}\rangle n_{\rho}(\tau) N_{K^*}(\tau) \nonumber \\
&+&\langle\sigma_{\rho\pi\to K^*\bar{K}}v_{\rho\pi} \rangle
n_{\pi}(\tau)N_{\rho}(\tau)-\langle \sigma_{K^*\bar{K}\to\rho
\pi}v_{K^*\bar{K}}\rangle n_{K}(\tau) N_{K^*}(\tau) \nonumber \\
&+& \langle\sigma_{\pi\pi\to K^*\bar{K}^*}v_{\pi\pi} \rangle
n_{\pi}(\tau)N_{\pi}(\tau)-\langle \sigma_{K^*\bar{K}^*\to
\pi\pi}v_{K^*\bar{K}^*}\rangle
n_{\bar{K}^*}(\tau) N_{K^*}(\tau) \nonumber \\
&+& \langle\sigma_{\rho\rho\to K^*\bar{K}^*}v_{\rho \rho} \rangle
n_{\rho}(\tau)N_{\rho}(\tau)-\langle \sigma_{K^*\bar{K}^*\to
\rho\rho}v_{K^*\bar{K}^*}\rangle
n_{\bar{K}^*}(\tau) N_{K^*}(\tau) \nonumber \\
&+&\langle \sigma_{\pi K\to K^*}v_{\pi K}\rangle n_{\pi}(\tau)
N_{K}(\tau)-\langle\Gamma_{K^*}\rangle N_{K^*}(\tau), \nonumber \\
\frac{dN_{K}(\tau)}{d\tau}&=&\langle\sigma_{\pi\pi\to
K\bar{K}}v_{\pi\pi} \rangle n_{\pi}(\tau)N_{\pi}(\tau)- \langle
\sigma_{K\bar{K}\to\pi\pi}v_{K\bar{K}}\rangle n_{\bar{K}}
(\tau)N_{K}(\tau)  \nonumber \\
&+& \langle\sigma_{\rho\rho\to K\bar{K}}v_{\rho\rho} \rangle
n_{\rho}(\tau)N_{\rho}(\tau)-\langle \sigma_{K\bar{K}\to
\rho\rho}v_{K\bar{K}}\rangle n_{\bar{K}}(\tau)
N_{K}(\tau) \nonumber \\
&+&\langle \sigma_{K^*\pi\to K\rho}v_{K^*\pi}\rangle n_{\pi}(\tau)
N_{K^*}(\tau)-\langle\sigma_{K\rho\to K^*\pi}v_{K\rho} \rangle
n_{\rho}(\tau)N_{K}(\tau)  \nonumber \\
&+&\langle \sigma_{K^*\rho\to K\pi}v_{K^*\rho}\rangle n_{\rho}
(\tau) N_{K^*}(\tau)-\langle\sigma_{K\pi\to K^*\rho}v_{K\pi}
\rangle n_{\pi}(\tau)N_{K}(\tau) \nonumber \\
&+&\langle\sigma_{\rho\pi\to K^*\bar{K}}v_{\rho\pi} \rangle
n_{\pi}(\tau)N_{\rho}(\tau)-\langle \sigma_{K^*\bar{K}\to
\rho\pi}v_{K^*\bar{K}}\rangle n_{\bar{K}}(\tau) N_{K^*}(\tau)
\nonumber \\
&+&\langle\Gamma_{K^*}\rangle N_{K^*}(\tau)-\langle \sigma_{\pi
K\to K^*}v_{\pi K}\rangle n_{\pi}(\tau) N_{K}(\tau),
\label{NKvKsrate}
\end{eqnarray} }
\end{widetext}
\noindent where $n_i(\tau)$ is the density of a light meson $i$ in
the hadronic matter at proper time $\tau$, and $N_j(\tau)$ is the
abundance of the other light meson $j$ in each process shown in
Fig. \ref{KvDiagrams} at proper time $\tau$. $n_i(\tau)$ for pions
and $\rho$ mesons is evaluated from

\begin{eqnarray}
&& n_i(\tau)=\frac{g_i}{2 \pi^2}\int_0^\infty
\frac{p^2dp}{e^{\sqrt{p^2+m_i^2}/T(\tau)}-1}
\nonumber \\
&& \quad\quad~\approx \frac{g_i}{2\pi^2}m_i^2 T(\tau)
K_2\Big(\frac{m_i}{T(\tau)}\Big), \label{Stat}
\end{eqnarray}
by assuming that they are in thermal equilibrium, and varies in
time through the temperature profile introduced below, Eq.
(\ref{Tprofiles}). We obtain $N_j(\tau)$ by multiplying Eq.
(\ref{Stat}) by the hadronization volume $V(\tau)$. In Eq.
(\ref{Stat}), $g_i$ is the degeneracy factor for a particle $i$
and $K_2$ the modified Bessel function of the second kind.

$n_i(\tau)$ and $N_j(\tau)$ are functions of the proper time
through the temperature profile developed to describe the dynamics
of relativistic heavy ion collisions. We use the schematic model
of a system with an accelerated transverse expansion based on the
boost invariant Bjorken picture \cite{Chen:2003tn, Chen:2007zp}.

\begin{eqnarray}
&& V(\tau)=\pi[R_c+v_c(\tau-\tau_c)+a_c/2(\tau-\tau_c)^2]^2\tau
c, \nonumber \\
&& T(\tau)=T_c-(T_h-T_f)\bigg(\frac{\tau-\tau_h}{\tau_f-\tau_h}
\bigg)^{4/5}, \label{Tprofiles}
\end{eqnarray}
with $T_h$ and $\tau_f$ being the hadronization temperature and
the freeze-out time, respectively. Eq. (\ref{Tprofiles}) describes
the system of the quark-gluon plasma expanding with its transverse
velocity $v_c$ and transverse acceleration $a_c$ starting from its
final transverse size $R_c$ at the chemical freeze-out time
$\tau_c$. The temperature of the system decreases from the
hadronization temperature to the kinetic freeze-out temperature
$T_f$. The values used in Eq. (\ref{Tprofiles}) are summarized in
Table \ref{data}.

\begin{table}[!h]
\caption{Values for the volume and temperature profiles in the
schematic model Eq. (\ref{Tprofiles}). } \label{data}
\begin{center}
\begin{tabular}{ccc}
\hline \hline
& Temp.(MeV) & Time (fm/c) \\
\hline
$R_c=8.0$ fm & $T_c=175$  & $\tau_c=5.0$  \\
$v_c=0.4 c$ & $T_h=175$  & $\tau_h=7.5$  \\
$a_c=0.02 c^2/$fm & $T_f=125$  & $\tau_f=17.3$  \\
\hline \hline
\end{tabular}
\end{center}
\end{table}

In the rate equations, Eq. (\ref{NKvKsrate}), $\left\langle
\sigma_{ab\rightarrow cd}v_{ab}\right\rangle$ is the thermally
averaged cross section for initial two particles in a two-body
process $ab\to cd$ given by \cite{Koch:1986ud}
{\allowdisplaybreaks
\begin{eqnarray} &&\left\langle \sigma
_{ab\rightarrow cd}v_{ab}\right\rangle
\nonumber \\
&&=\frac{1}{1+\delta_{ab}}\frac{\int d^{3}\vec p_{a}d^{3}\vec
p_{b}f_{a} (\vec p_{a})f_{b}(\vec p_{b})\sigma_{ab\to
cd}v_{ab}}{\int d^{3}\vec p_{a}d^{3}\vec p_{b}f_{a}(\vec p_{a})
f_{b}(\vec p_{b})} \nonumber\\
&&=\frac{1}{1+\delta_{ab}}\frac{T^4}{4m_a^2 K_2(m_a/T)m_b^2
K_2(m_b/T)} \nonumber \\
&& \times \int^\infty_{z_0}dzK_1(z)\sigma(z^2 T^2)
\nonumber \\
&&\times[z^2-(m_a+m_b)^2/T^2][z^2-(m_a-m_b)^2/T^2], \label{sigmav}
\end{eqnarray}}
with $z_0=\mathrm{max}((m_a+m_b)/T, (m_c+m_d)/T)$, $K_1$ and $K_2$
being the modified Bessel function of the second kind, $f_i$ being
the Boltzmann momentum distribution of the particle $i$, $f_i(\vec
p)=e^{-\sqrt{\vec p^2+m_i^2}}$, respectively. $v_{ab}$ is the
relative velocity of interacting particles $a$ and $b$,
$v_{ab}=\sqrt{(p_a\cdot p_b)^2-m^2_am^2_b}/(E_aE_b)$.
$\langle\Gamma_{K^*}\rangle$ in Eq. (\ref{NKvKsrate}) is the
thermally averaged decay width of $K^*$ mesons, Eq.
(\ref{totalG}),
$\langle\Gamma_{K^*}\rangle=\Gamma_{K^*}(m_{K^*})K_1(m_{K^*}/T)/
K_2(m_{K^*}/T)$, which has been obtained in the same methods as
used in Eq. (\ref{sigmav}).

\begin{widetext}

\begin{figure}[!h]
\begin{center}
\includegraphics[width=0.495\textwidth]{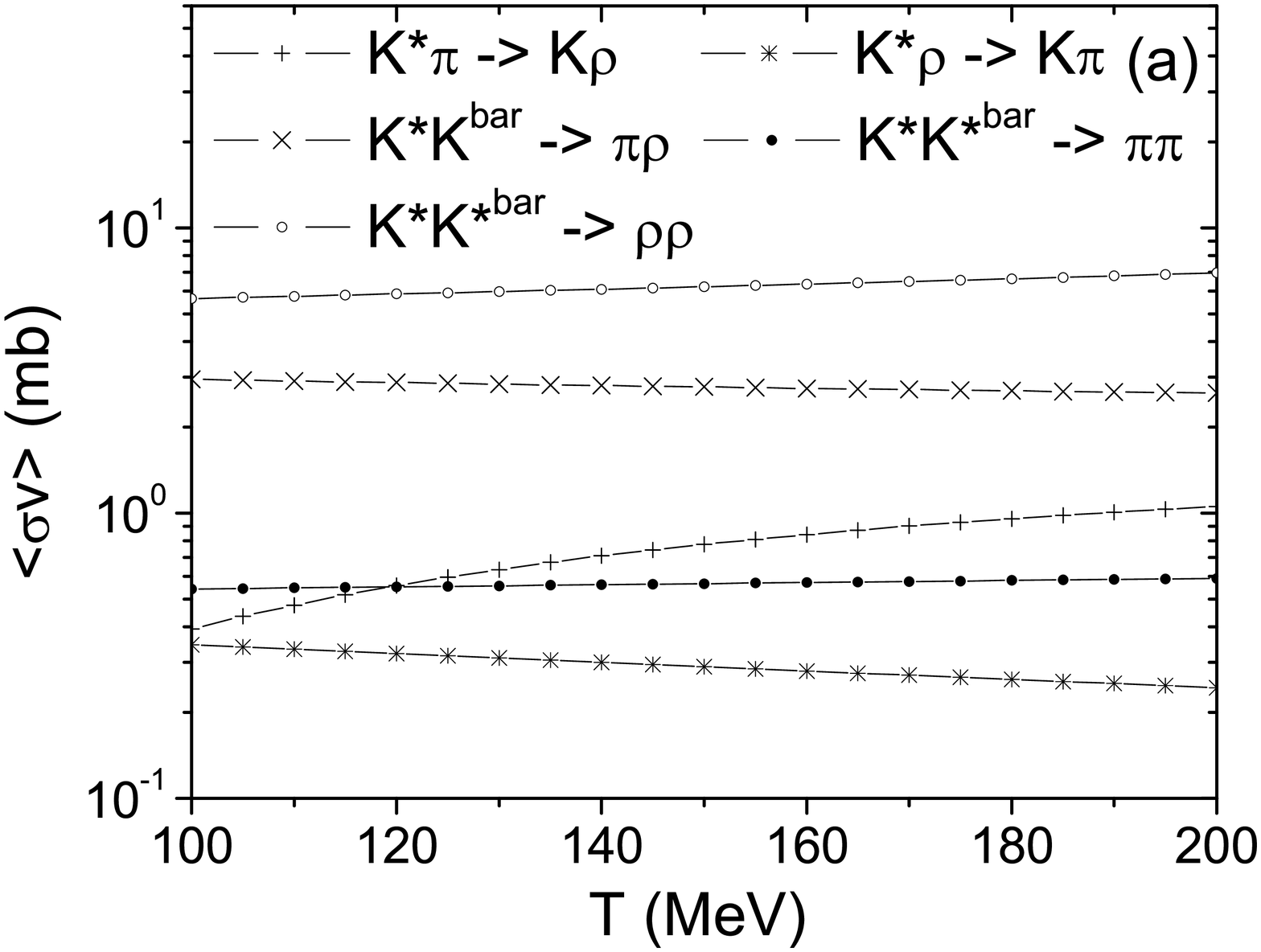}
\includegraphics[width=0.495\textwidth]{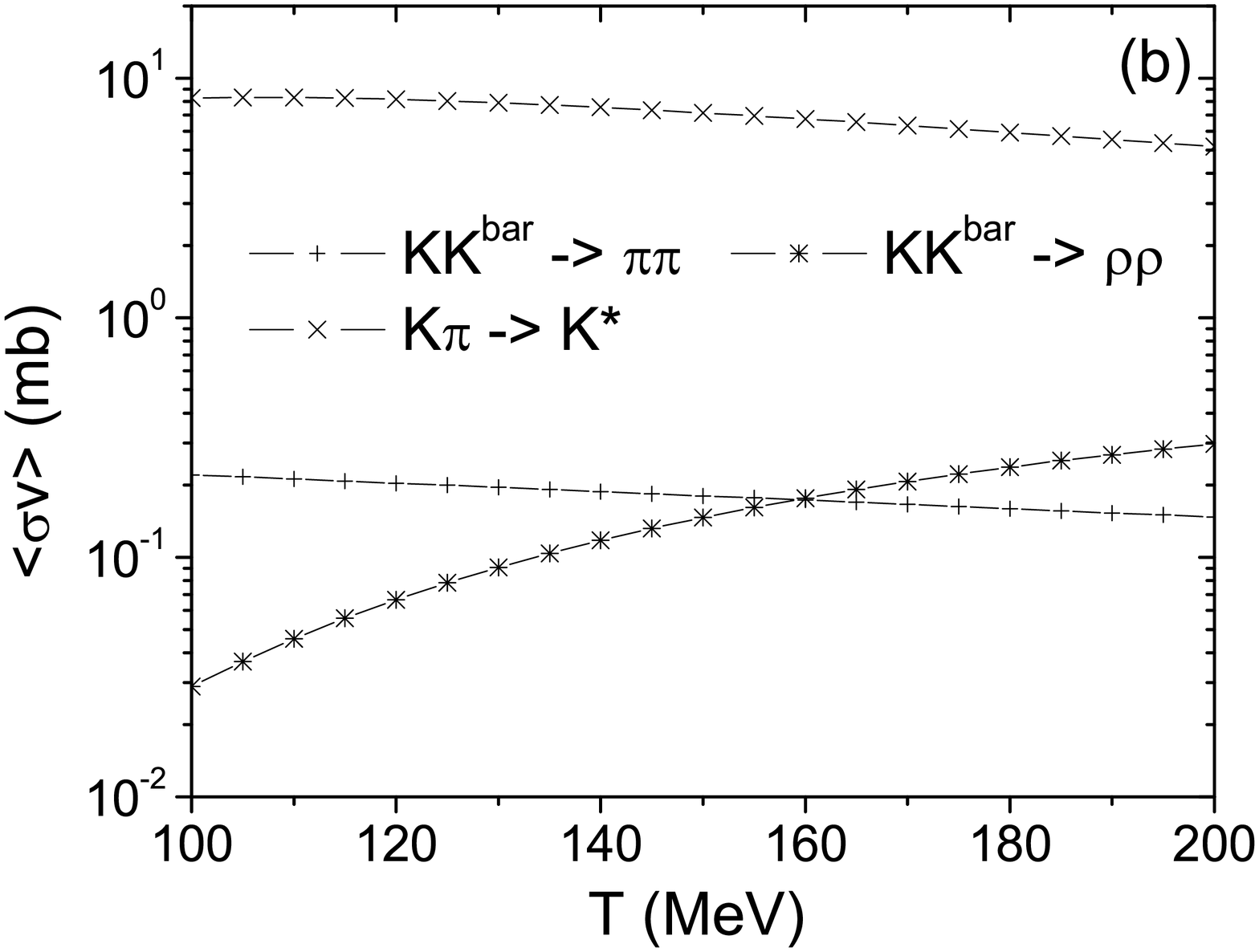}
\includegraphics[width=0.495\textwidth]{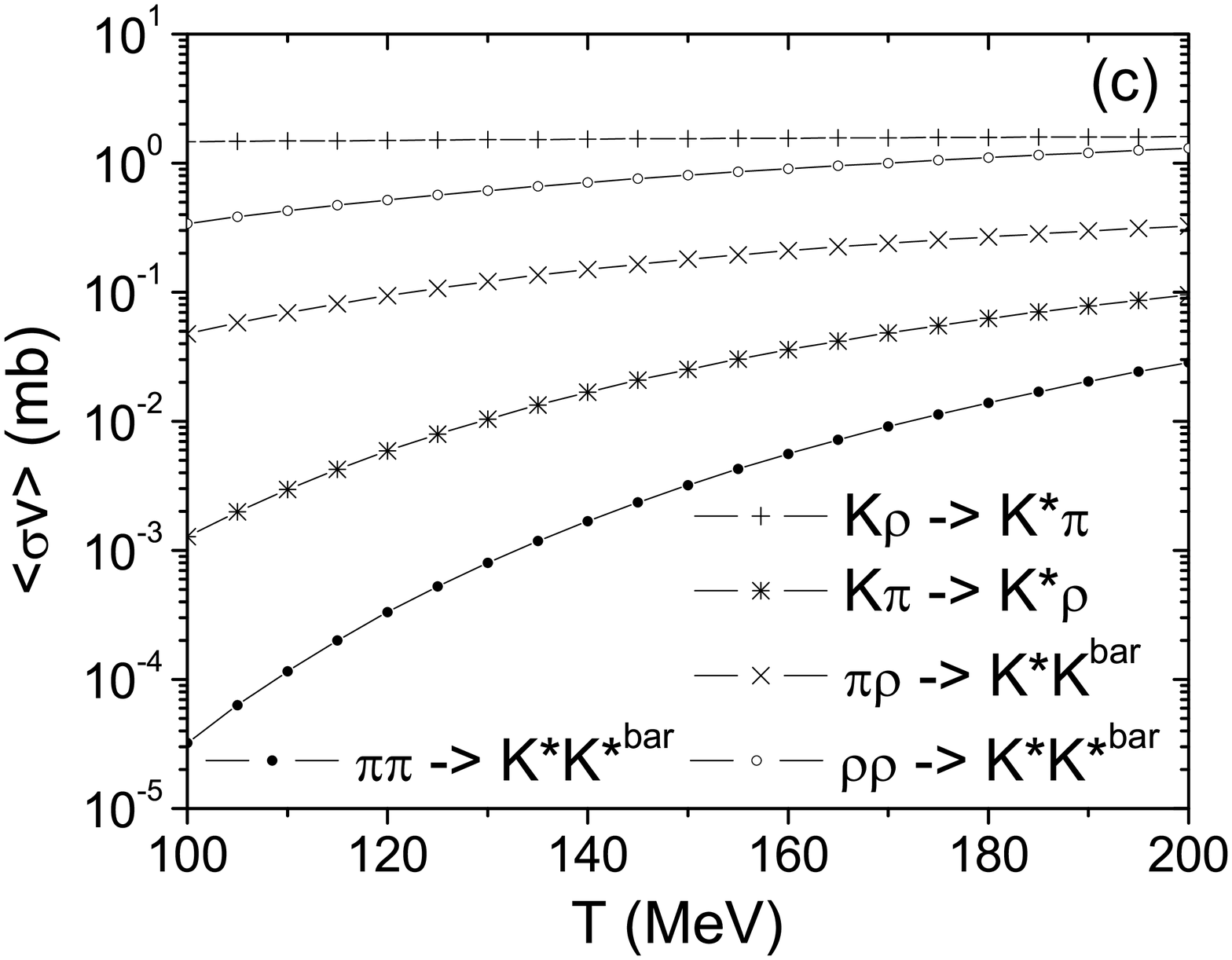}
\includegraphics[width=0.495\textwidth]{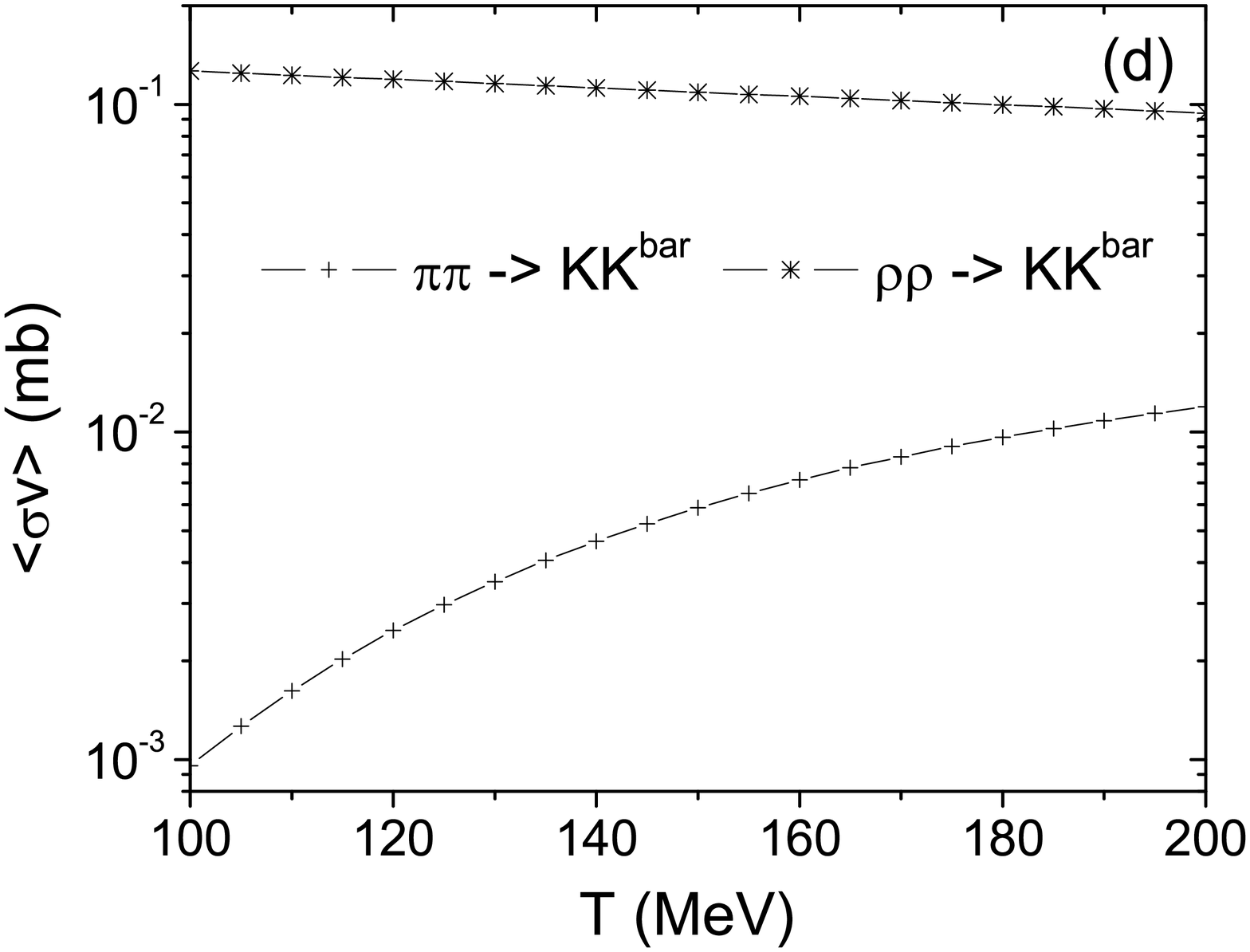}
\end{center}
\caption{Thermally averaged cross sections for the absorption of
(a) a $K^*$ meson by pions, $\rho$, $K$, and $K^*$ mesons via
processes $K^*\pi\rightarrow\rho K$, $K^*\rho\rightarrow \pi K$,
$K^*\bar{K}\rightarrow \rho\pi$, $K^*\bar{K}^*\rightarrow \pi\pi$,
and $K^*\bar{K}^*\rightarrow \rho\rho$, and those for (b) a $K$
meson via processes $K\bar{K}\rightarrow \pi\pi$,
$K\bar{K}\rightarrow \rho\rho$, and $K\pi\rightarrow K^*$.
Thermally averaged cross sections for their inverse processes
$\rho K\rightarrow K^*\pi$, $\pi K\rightarrow K^*\rho$,
$\rho\pi\rightarrow K^*\bar{K}$, $\pi\pi\rightarrow K^*\bar{K}^*$,
and $\rho\rho\rightarrow K^*\bar{K}^*$ for (c) a $K^*$ meson,
$\pi\pi\rightarrow K\bar{K}$ and $\rho\rho\rightarrow K\bar{K}$
for (d) a $K$ meson.} \label{KvKssigmav}
\end{figure}
\end{widetext}

The $K^*$ meson abundance at $\tau$, $N_{K^*}$, depends not only
on the dissociation reactions like $K^*\pi\rightarrow\rho K$,
$K^*\rho\rightarrow \pi K$, $K^*\bar{K}\rightarrow \rho\pi$,
$K^*\bar{K}^*\rightarrow \pi\pi$, and $K^*\bar{K}^*\rightarrow
\rho\rho$, but also on the production reactions, or the inverse
reactions of the dissociation reactions, such as $\rho
K\rightarrow K^*\pi$, $\pi K\rightarrow K^*\rho$,
$\rho\pi\rightarrow K^*\bar{K}$, $\pi\pi\rightarrow K^*\bar{K}^*$,
and $\rho\rho\rightarrow K^*\bar{K}^*$. We have taken both
reactions into consideration in building the coupled equation for
both the $K^*$ meson and kaon in Eq. (\ref{NKvKsrate}). We have
used the detailed balance relation when evaluating thermally
averaged cross sections of the inverse reactions from the results
for forward processes shown in Fig. \ref{KvKssigma}. The results
are shown in Fig. \ref{KvKssigmav}.

As we see in Fig. \ref{KvKssigmav}, thermally averaged cross
sections of the dissociation reactions are bigger than those of
the production reactions for the exothermic reactions. In the case
of the endothermic reaction like $K^*\pi\rightarrow\rho K$,
thermalized production cross section is bigger than that for
dissociation reaction. Both themalized cross sections are
comparable for the other endothermic reaction,
$K\bar{K}\rightarrow \rho\rho$.

In general, the smaller threshold energy, mass, and degeneracy,
the bigger thermally averaged cross section in the two-body
process. We find that the thermally averaged cross section for
$K^*$ formation, $K\pi\to K^*$ becomes more significant than those
for other reactions. However, the unusually rising cross section
in energy for the reaction $K^*\bar{K}^*\rightarrow \rho\rho$ has
been suppressed in the thermalized medium as shown in Fig.
\ref{KvKssigmav}(a).

When solving the coupled differential equation for both the $K^*$
meson and kaon abundances, we have treated abundances of their
antiparticles $\bar{K}^*$ and $\bar{K}$ mesons also as variables
using the strangeness chemical potential $\mu_s$; i.e.,
$N_{\bar{K}^*}=e^{-2\mu_s/T(\tau)}N_{K^*}$ and same for antikaons.
In other words, we have not considered that $K^*$ mesons and kaons
are in thermal equilibrium during the expansion of the hadronic
matter, while we calculate the thermally averaged cross section
Eq. (\ref{sigmav}) using the thermal distributions of hadrons
involved. However, the initial yield of kaons at chemical
freeze-out has been evaluated to be 88.1 using the statistical
hadronization model, Eq. (\ref{Stat}) with the strangeness
chemical potential $\mu_s=10$ MeV and the hadronization volume
$V_H=1908$ fm$^3$ \cite{Cho:2013rpa}, whereas the initial yield of
$K^*$ mesons has been obtained from,
\begin{eqnarray}
&& N_{K^*}(\tau)=V_H\frac{g_{K^*}}{2 \pi^2}\int_{m_{th}}^\infty
\frac{dm}{N_{BW}}\frac{\Gamma_{K^*}}{(m-m_{K^*})^2+\Gamma_{K^*}^2/4}
\nonumber \\
&& \qquad \times \int_0^\infty\frac{p^2dp}{e^{(\sqrt{p^2+m_{K^*}^2
}-\mu_s)/T(\tau)}-1}, \label{StatWidth}
\end{eqnarray}
to take the width of the $K^*$ meson into consideration. In Eq.
(\ref{StatWidth}), $m_{th}$ is the threshold energy for $K^*\to
K\pi$ decay channel and $N_{BW}$ is the normalization constant for
the Breit-Wigner distribution. We obtain the $K^*$ meson initial
yield to be 55.7, which is slightly larger than 52.4 obtained
without including the $K^*$ meson width calculated with the
formula given in  Eq. (\ref{Stat}).

\begin{figure}[!t]
\begin{center}
\includegraphics[width=0.53\textwidth]{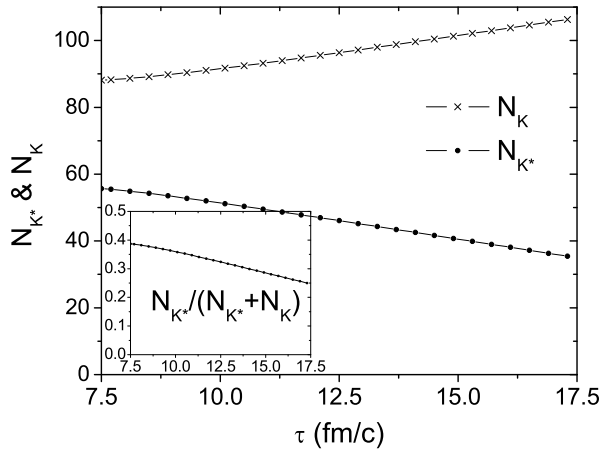}
\end{center}
\caption{Time evolution of the $K^*$ meson and kaon abundances
during the hadronic stage in central Au+Au collisions at
$\sqrt{s_{NN}}$ = 200 GeV. Ratio of the $K^*$ meson abundance to
the sum of the $K^*$ meson and kaon abundances is shown in the
inset. } \label{KvKsEvol}
\end{figure}

In Fig. \ref{KvKsEvol}, we show the abundances of the $K^*$ meson
and kaon as a function of the proper time during the hadronic
stage of heavy ion collisions at $\sqrt{s_{NN}}$ = 200 GeV. As we
have expected, the $K^*$ meson abundance decreases due to both
interactions of $K^*$ mesons with other hadrons and the decay of
the $K^*$ meson to the pion and kaon, eventually becoming 35.6 at
9.8 fm/s after the chemical freeze-out. On the other hand the
abundance of the kaon increases up to 106.5 at the end of hadronic
expansion. We find that throughout the time evolution the sum of
the $K^*$ meson and kaon abundances changes slightly from 143.8 to
142.1. We also show in the inset of Fig. \ref{KvKsEvol} the
variation of the ratio of the $K^*$ meson abundance to the sum of
the $K^*$ meson and kaon abundances. The ratio decreases from the
initial ratio from the statistical hadronization model 0.39 to
0.25 in the end.

Based on the analysis we find that about $36\%$ of $K^*$ mesons
produced at chemical freeze-out disappear during the hadronic
stage in heavy ion collisions, making the invariant mass
reconstruction of the total $K^*$ meson difficult. We further find
that the hadronic interactions shown in both Fig. \ref{KvDiagrams}
and Fig. \ref{KDiagrams} explain about $6\%$ of the $K^*$ meson
loss, and the $K^*$ meson decay is largely responsible for the
$K^*$ meson reduction in the hadronic medium. Our result is
comparable to the $30\%$ reduction of the previous statistical
model prediction $0.33 \pm 0.01$ \cite{BraunMunzinger:2001ip} to
the experimental measurements $0.23 \pm 0.05$ \cite{Adams:2004ep}.

\begin{figure}[!t]
\begin{center}
\includegraphics[width=0.53\textwidth]{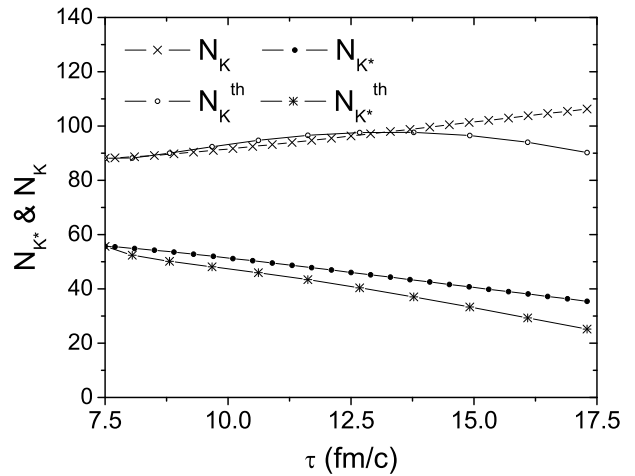}
\end{center}
\caption{A comparison between the $K^*$ meson and kaon abundances
due to all hadronic interactions shown in Fig. \ref{KvDiagrams}
and Fig. \ref{KDiagrams} and those from the thermal model
prediction at each time and temperature.} \label{KvKsEvolthS}
\end{figure}

We also consider the possibility of both the $K^*$ meson and kaon
thermalization during the hadronic expansion. Assuming that both
mesons are in thermal equilibrium with the hadronic medium we
evaluate the $K^*$ meson and kaon abundances in time using Eqs.
(\ref{Stat}), (\ref{StatWidth}), and (\ref{Tprofiles}), and show
the results represented by $N_K^{th}$ and $N_{K^*}^{th}$ in Fig.
\ref{KvKsEvolthS}.

As we see, $N_{K^*}^{th}$ keeps decreasing all the time. However,
$N_K^{th}$ increases at the beginning of the hadronic stage, and
finally decreases. This is due to the competition between the
volume expansion and the decreasing rate caused by the thermal
effects in Eq. (\ref{Stat}), through the factor $m_K/T(\tau)$
inside the modified Bessel function of the second kind, implying
that $N_K^{th}$ and $N_{K^*}^{th}$ depends on the size and also on
the lifetime of the expanding fireball, Eq. (\ref{Tprofiles}).
Nevertheless, one should note that the volume of the system
expands in time with the total entropy almost preserved
\cite{Siemens:1979dz}. We find that abundances of most hadrons
decrease during the hadronic expansion in the statistical
hadronization model but that of the pion, the lightest hadron,
increases in time to compensate the loss of entropy from heavier
hadrons. We argue that the same mechanism is happening to
strangeness hadrons for some time during the hadronic expansion.
However, the ratio $N_{K^*}^{th}/(N_K^{th}+N_{K^*}^{th})$ is not
affected by the volume, and it keeps decreasing from 0.39 to 0.22
at the kinetic freeze-out.

We notice that recent measurements of the $K^*$ meson yield in
Pb+Pb collisions at $\sqrt{s_{NN}}=2.76$ TeV at the Large Hadron
Collider (LHC) \cite{Abelev:2014uua} provide $0.19 \pm 0.05$ as
the $K^{*0}/K^-$ ratio. This value is also smaller than the
statistical hadronization model prediction 0.30 evaluated with the
hadronization temperature 156 MeV \cite{Stachel:2013zma} at the
LHC energy. The measurements indicate that more $K^*$ mesons are
lost during the hadronic expansion at LHC, leading to $37\%$
reduction of the ratio.

\section{The abundance ratio of $K^*$ mesons to kaons in heavy
ion collisions}

Since the interactions of $K^*$ mesons and kaons with light mesons
considered in Fig. \ref{KvDiagrams} and Fig. \ref{KDiagrams} take
place during the hadronic stage at both RHIC and LHC, it is
necessary to understand general features of the variation of the
$K^*$ meson abundance in heavy ion collisions. In order to analyze
the reduction of the $K^*$ meson in the hadronic medium we
simplify the coupled equation, Eq. (\ref{NKvKsrate}) by keeping
the linear terms in $ N_K$ and $N_{K^*}$ only.
\begin{eqnarray}
&&\frac{dN_{K^*}(\tau)}{d\tau}=\gamma_K N_{K}(\tau)-\gamma_{K^*}
N_{K^*}(\tau), \nonumber \\
&&\frac{dN_K(\tau)}{d\tau}=-\gamma_K
N_{K}(\tau)+\gamma_{K^*}N_{K^*}(\tau), \label{NKvKsrateSimple}
\end{eqnarray}
with
\begin{eqnarray}
&& \gamma_{K^*}=\langle\sigma_{K^*\rho\to K\pi}v_{K^*\rho} \rangle
n_{\rho}+\langle\sigma_{K^*\pi\to K\rho}v_{K^*\pi} \rangle n_{\pi}
\nonumber \\
&& \quad~~+\langle\Gamma_{K^*}\rangle, \nonumber \\
&& \gamma_K=\langle\sigma_{K\pi\to K^*\rho}v_{K\pi} \rangle
n_{\pi}+\langle\sigma_{K\rho\to K^*\pi}v_{K\rho} \rangle n_{\rho}
\nonumber \\
&& \quad~+\langle\sigma_{K\pi\to K^*}v_{K\pi}\rangle n_{\pi}.
\label{gammas}
\end{eqnarray}
When the thermal cross sections and densities of light mesons are
independent of time, following analytic solutions are obtained
from the coupled equation, Eq. (\ref{NKvKsrateSimple}),
\begin{eqnarray}
&& N_{K^*}(\tau)=\frac{\gamma_K}{\gamma}N^0+\Big(N_{K^*}^0-
\frac{\gamma_K}{\gamma}N^0\Big)e^{-\gamma(\tau-\tau_h)},
\nonumber \\
&& N_K(\tau)=\frac{\gamma_{K^*}}{\gamma}N^0+\Big(N_K^0-
\frac{\gamma_{K^*}}{\gamma}N^0\Big)e^{-\gamma(\tau-\tau_h)},
\label{AnalSol}
\end{eqnarray}
where the initial yields for both hadrons, $N_K^0$ and $N_{K^*}^0$
have been assumed, and $N^0$ is the sum of the $K^*$ meson and
kaon yields, $N^0=N_K^0+N_{K^*}^0$ at chemical freeze-out. The
$\gamma$ in Eq. (\ref{AnalSol}) is the sum of the $K^*$ meson and
kaon widths in the hadronic phase, $\gamma=\gamma_K+\gamma_{K^*}$;
$\gamma_{K^*}$ and $\gamma_K$ play roles of the collisional
broadening of the width of the $K^*$ meson and kaon in the
hadronic medium, respectively.

In Eq. (\ref{AnalSol}) time-independent terms represent abundances
when time goes to infinity, and the sum of two solutions is
preserved as its initial value $N_K^0+N_{K^*}^0$. As time goes on
$N_K$ increases while $N_{K^*}$ decreases, and the rate at which
the final number is reached in Eq. (\ref{AnalSol}) is determined
by the $\gamma$ which takes into account hadronic interactions of
$K^*$ mesons and kaons with light mesons. If the $\gamma$ is
large, the abundance can change significantly for a short time.

Let us now investigate the time evolution of yield ratio of $K^*$
mesons to kaons from the analytic solution of Eq. (\ref{AnalSol}),
$R(\tau)=N_{K^*}(\tau)/(N_K^*(\tau)+N_K(\tau))$;
\begin{eqnarray}
&& R(\tau)=\frac{N_{K^*}(\tau)}{N_{K^*}(\tau)+N_K(\tau)}=
\frac{N_{K^*}(\tau)}{N^0} \nonumber \\
&& \qquad~= \frac{\gamma_K}{\gamma}+\Big(\frac{N_{K^*}^0}{N^0}-
\frac{\gamma_K}{\gamma}\Big)e^{-\gamma(\tau-\tau_h)}.
\label{ratio}
\end{eqnarray}

We notice that $R(\tau)$ is also composed of two parts; a
time-independent part and a transient part. After a long time
$\tau$ the time-independent part $R(\infty)=\gamma_K/\gamma$ is
expected to represent the $K^*$ meson to the kaon ratio. How fast
the abundance ratio approaches the time-independent part relies on
$\gamma$, the sum of the $K^*$ meson and kaon widths in the
exponential function.

\begin{figure}[!t]
\begin{center}
\includegraphics[width=0.52\textwidth]{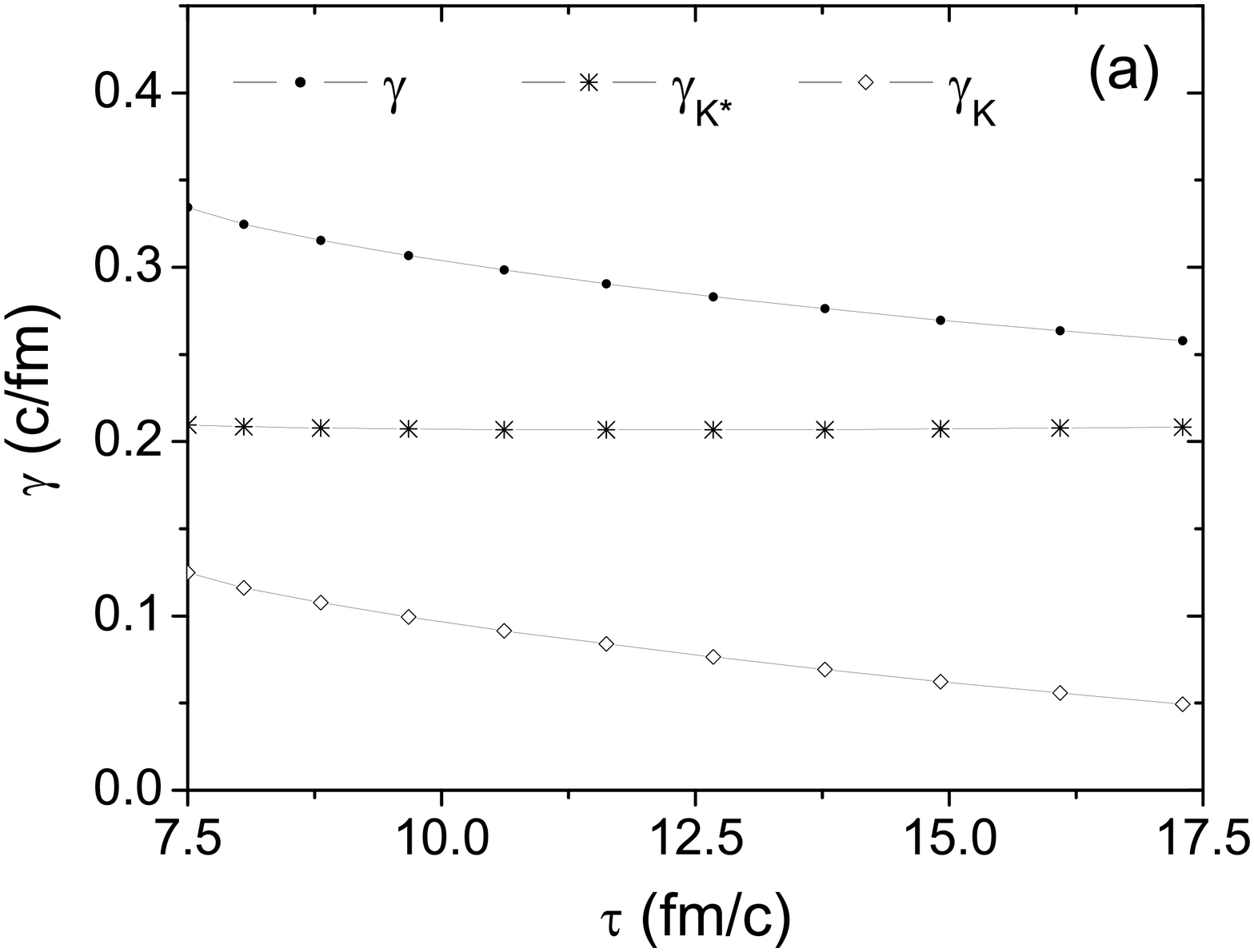}
\includegraphics[width=0.51\textwidth]{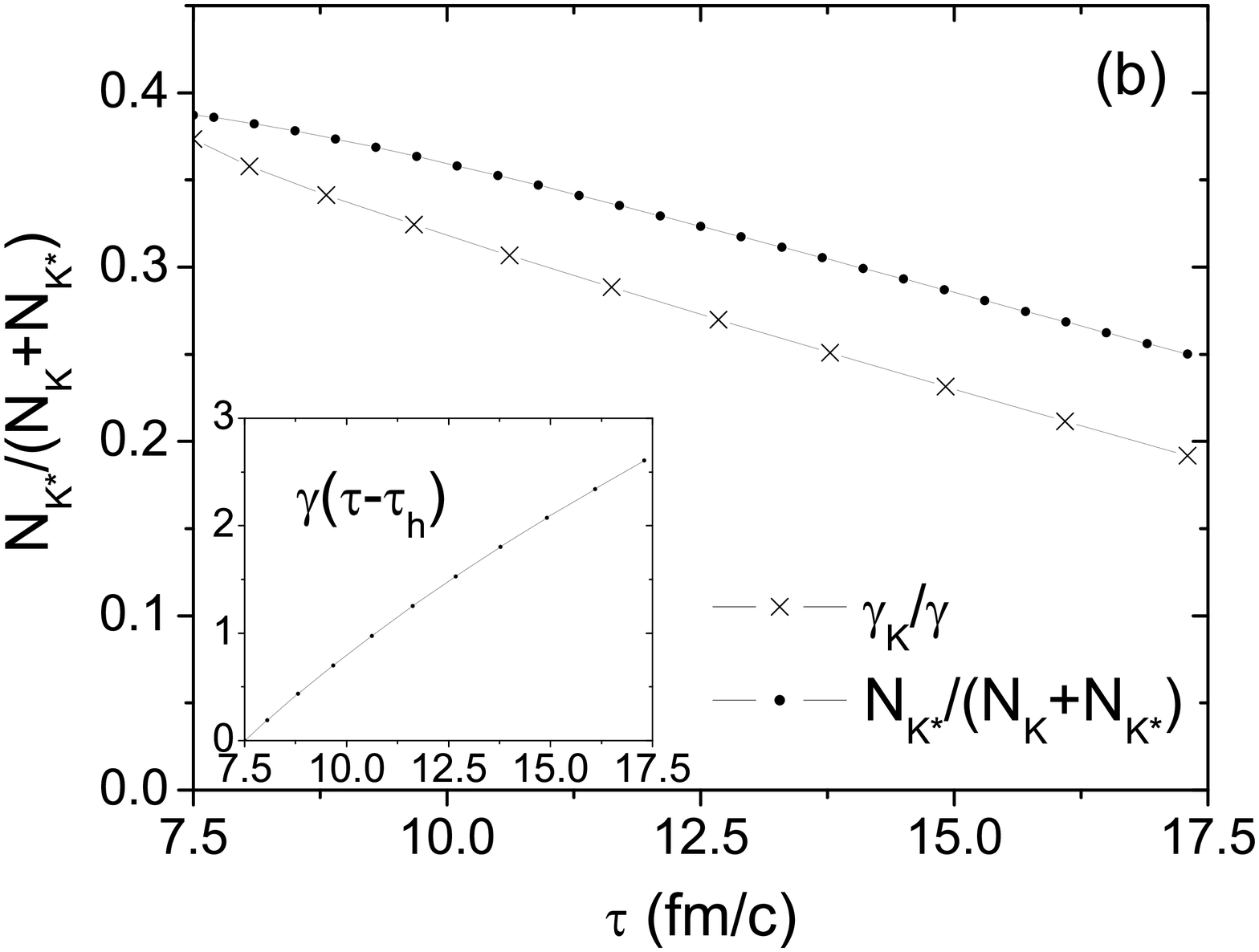}
\end{center}
\caption{(a) Variations of $\gamma_K$, $\gamma_{K^*}$, and
$\gamma=\gamma_K+\gamma_{K^*}$ in time during the hadronic stage.
(b) A comparison of the abundance ratio variation
$N_{K^*}/(N_K+N_{K^*})$ obtained numerically from Eq.
(\ref{NKvKsrate}) for RHIC and $R(\tau=\infty)=\gamma_K/\gamma$
from Eq. (\ref{ratio}) evaluated at each time and temperature. We
show in the inset how $\gamma(\tau-\tau_h)$ changes in time.}
\label{gamma_ratio}
\end{figure}

With these in mind let us investigate the variation in the
abundance of $K^*$ mesons and kaons obtained from Eq.
(\ref{NKvKsrate}). Since all thermally averaged cross sections and
densities of the light mesons in $\gamma_K$ and $\gamma_{K^*}$ are
functions of a time, solutions of Eq. (\ref{NKvKsrate}) are
different from the analytic solution of the simplified equation,
Eq. (\ref{NKvKsrateSimple}). Nevertheless, we find that the
solution of Eq. (\ref{NKvKsrate}) keeps the same important
characteristics of the analytic solutions from Eq.
(\ref{NKvKsrateSimple}).

We first show in Fig. \ref{gamma_ratio}(a) $\gamma$'s obtained at
Eq. (\ref{NKvKsrate}) as a function of time. The $\gamma$
decreases in time from 0.33 c/fm to 0.26 c/fm as the system cools
down from 175 MeV at $\tau_H$ to 125 MeV at $\tau_f$, reflecting
that interactions between hadrons become less vigorous as the
temperature of the system decreases. The $\gamma_K$, or
$\langle\sigma_{K\pi\to K^*\rho}v_{K\pi} \rangle
n_{\pi}+\langle\sigma_{K\rho\to K^*\pi}v_{K\rho} \rangle n_{\rho}
+\langle\sigma_{K\pi\to K^*}v_{K\pi}\rangle n_{\pi}$ in Eq.
(\ref{gammas}) decreases as the temperature decreases, but
$\gamma_{K^*}$ is almost constant during the hadronic stage; with
decreasing temperature of the system, the part of $\gamma_{K^*}$,
or $\langle\sigma_{K^*\rho\to K\pi}v_{K^*\rho} \rangle
n_{\rho}+\langle\sigma_{K^*\pi\to K\rho}v_{K^*\pi} \rangle
n_{\pi}$ decreases whereas the thermal width of the $K^*$ meson
$\langle\Gamma_{K^*}\rangle$ slightly increases due to the factor
$K_1(m_{K^*}/T)/K_2(m_{K^*}/T)$, meaning that $K^*$ mesons live
shorter at lower temperature.

We compare in Fig. \ref{gamma_ratio}(b) the abundance ratio
variation $N_{K^*}/(N_K+N_{K^*})$ evaluated numerically from Eq.
(\ref{NKvKsrate}) for RHIC to $\gamma_K/\gamma$ obtained from Eq.
(\ref{ratio}) at each time and temperature. $\gamma_K/\gamma$
represents the expected hadronic interaction width ratio between
kaons and $K^*$ mesons plus kaons at each temperature and time. We
anticipate that the abundance ratio of $K^*$ mesons and kaons in
Fig. \ref{gamma_ratio} approaches to $\gamma_K/\gamma$ as time
goes on, like the ratio between those mesons obtained from the
simplified rate equation, Eq. (\ref{ratio}). We show in the inset
of Fig. \ref{gamma_ratio}(b) how $\gamma(\tau-\tau_h)$ varies in
time at RHIC. As we see, $\gamma(\tau-\tau_h)$ increases up to 2.5
for 9.8 fm/c. We expect that the similar term with
$e^{-\gamma(\tau-\tau_h)}$ in the real solution suppresses the
contribution of the time-dependent term as time goes on. The
discrepancy between the abundance ratio and the $\gamma_K/\gamma$
in Fig. \ref{gamma_ratio}(b) is attributable to both the
contribution from non-linear terms included in Eq.
(\ref{NKvKsrate}) such as $K^*\bar{K}\to \rho\pi$, (e) and (f);
$K^*\bar{K}^*\to \pi\pi$, (g) and (h); $K^*\bar{K}^*\to \rho\rho$,
(i) and (j) shown in Fig. \ref{KvDiagrams} and
$K\bar{K}\to\pi\pi$, (a) and (b); $K\bar{K}\to\rho\rho$, (c) and
(d) shown in Fig. \ref{KDiagrams}, and the time delay required to
reach thermal equilibrium from the interactions of $K^*$ mesons
with light mesons in the hadronic medium.

Based on the above analysis we argue that the final ratio of the
yield between $K^*$ mesons and kaons in heavy ion collisions is
largely dependent on their interactions with other hadrons in the
hadronic medium, $\gamma_K$ and $\gamma_{K^*}$. Since
$\gamma(\tau-\tau_h)$ keeps increasing during the hadronic stage,
the transient term with $e^{-\gamma(\tau-\tau_h)}$ in the yield
ratio $R(\tau)$ plays a negligible role at a later time during the
hadronic interaction stage. Therefore, we see that the relative
interaction ratio $\gamma_K/\gamma$ mainly determines the final
yield ratio between $K^*$ mesons and kaons at the end of the
hadronic stage.

\begin{figure}[!t]
\begin{center}
\includegraphics[width=0.50\textwidth]{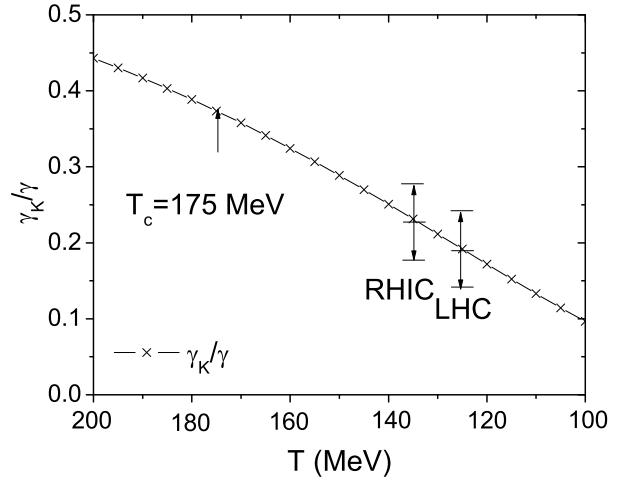}
\end{center}
\caption{A variation of the $\gamma_K/\gamma$ in temperature
during the hadronic stage. We show measurements of the abundance
ratio between $K^*$ mesons and kaons, $N_{K^*}/(N_K+N_{K^*})$
0.23$\pm$0.05 at RHIC \cite{Adams:2004ep} and 0.19$\pm$0.05 at LHC
\cite{Abelev:2014uua} } \label{gamKtogam}
\end{figure}
We show in Fig. \ref{gamKtogam} $\gamma_K/\gamma$ as a function of
the temperature of the system. We also show in Fig.
\ref{gamKtogam} measurements of the abundance ratio between $K^*$
mesons and kaons, $N_{K^*}/(N_K+N_{K^*})$, 0.23$\pm$0.05 at RHIC
\cite{Adams:2004ep} and 0.19$\pm$0.05 at LHC
\cite{Abelev:2014uua}. We notice from Fig. \ref{gamKtogam} that
the ratio of the $K^*$ meson and kaon in heavy ion collisions
seems to reflect the interaction ratio between strange and light
mesons, $\gamma_K/\gamma$ at the kinetic freeze-out temperature.
We infer that the lower ratio $N_{K^*}/(N_K+N_{K^*})$ at LHC
compared to that at RHIC is due to a lower kinetic freeze-out
temperature at LHC than at RHIC.

It has been argued that the degree of the reduction of $K^*$ meson
yield during the hadronic stage in heavy ion collisions is
attributable to a lifetime of the hadronic stage. Since the system
of quark-gluon plasma at LHC is much larger than that at RHIC, it
has been assumed that the lifespan of the hadronic stage at LHC is
also longer compared to that at RHIC, and thereby more $K^*$ meson
are lost in the hadronic medium at LHC.

We find, however, from the investigation of the variation in the
yield ratio between $K^*$ mesons and kaons based on the solution
of the rate equation, that the reduction of the $K^*$ meson in
heavy ion collisions reflects the interaction of $K^*$ mesons and
kaons with light mesons at kinetic freeze-out. We argue that the
degree of the $K^*$ meson abundance reduction in heavy ion
collisions, or the reduction of the yield ratio between the $K^*$
meson and kaon, is largely attributable to \textit{the kinetic
freeze-out temperature} via the interaction of $K^*$ mesons and
kaons with light mesons in the hadronic medium. The long lifespan
of the hadronic stage just suppresses more a transient term, such
as the second term in Eq. (\ref{ratio}), contributing little to
the change of the $K^*$ meson to kaon ratio.

As has been already mentioned, how fast meson abundances change in
the hadronic medium is governed by the sum of all the interactions
involved, i.e., the width $\gamma$ in Eq. (\ref{AnalSol}).
Therefore, in addition to the hadronic interactions considered in
Figs. \ref{KvDiagrams} and \ref{KDiagrams} all other interactions
with various hadrons, i.e., nucleons, have to be taken into
account to thoroughly understand the reduction of $K^*$ mesons in
heavy ion collisions. Moreover, we also have to include more the
feed down effects from heavier strangeness hadrons to fully
consider the abundance ratio of the $K^*$ meson to the kaon, which
are left for the future work.

\section{Conclusion}

We have studied the reduction of $K^*$ mesons in heavy ion
collisions. We have focused on the hadronic effects on the $K^*$
meson and kaon abundances during the hadronic stage of the cental
Au+Au collisions at $\sqrt{s_{NN}}$ = 200 GeV in order to
understand the $K^*$ meson yield difference between the
experimental measurement and the statistical hadronization model
prediction. We have evaluated absorption cross sections for both
kaons and $K^*$ mesons by pions, $\rho$, $K$, and $K^*$ mesons
inside the hadronic medium. In describing the interaction between
$K^*$ mesons and kaons and light mesons, we have introduced one
meson exchange model with the effective Lagrangian. Furthermore,
we have built the coupled differential equation for $K^*$ mesons
and kaons, and have solved it to investigate the time evolution of
the $K^*$ meson and kaon abundances during the expansion of the
hadronic matter.

We have found that the $K^*$ meson and kaon abundances during the
hadronic stage of heavy ion collisions are dependent on absorption
cross sections and their thermal average. We have shown that the
sum of $K^*$ and kaon abundances are almost preserved during the
expansion, and the interaction of $K^*$ mesons with light mesons
controls the reduction or production of $K^*$ mesons and kaons in
the hadronic matter. Our analysis indicates that $36\%$ of the
total $K^*$ mesons produced at the chemical freeze-out are lost
during the hadronic expansion in the central Au+Au collisions at
$\sqrt{s_{NN}}=200$ GeV. We have found that among $36\%$ about
$6\%$ of the total $K^*$ mesons are converted into light mesons by
hadronic interactions, and the remaining $30\%$ reduction is due
to the decay of $K^*$ mesons to kaons and pions. We see that the
loss of the $K^*$ meson abundance in the hadronic medium explains
very well the discrepancy of the $K/K^*$ ratio between the
statistical hadronization \cite{BraunMunzinger:2001ip} model
prediction and the experimental measurements \cite{Adams:2004ep}.

We have shown that the results obtained here can be applied to the
analysis of the $K^*$ meson production at the LHC. We have found
that all the interactions involved at RHIC must be present at LHC,
and therefore widths $\gamma_{K^*}$ and $\gamma_K$ evaluated at
the RHIC energy can be applied to the case at the LHC energy.
Moreover, we have realized that the smaller ratio of $K^*/K$
measured at the LHC energy indicates a lower temperature of the
kinetic freeze-out at LHC compared to that at RHIC. We have shown
that the yield ratio between $K^*$ mesons and kaons is not mainly
dependent on the lifetime of the hadronic stage in heavy ion
collisions, and the hadronic interaction width ratio of strange
mesons with light mesons, $\gamma_K/\gamma$ determines the final
yield ratio between $K^*$ mesons and kaons. We therefore conclude
that studying the yield of the $K^*$ meson and its variation
during the hadronic stage in relativistic heavy ion collisions
provides a chance to understand not only the production of $K^*$
mesons but also the evolution of the hadronic medium in heavy ion
collisions.

\section*{Acknowledgements}

S. Cho was supported by 2015 Research Grant from Kangwon National
University. S. H. Lee was supported by the Korea National Research
Foundation under the grant number KRF-2011-0020333 and
KRF-2011-0030621.

\end{document}